\documentstyle[amssymb,preprint,aps]{revtex}

\begin{document}
\draft
\title{Consistent Quantization of Noncommutative D-brane in Non-constant B Field
Background}
\author{Jen-Chi Lee\thanks{
e-mail: jcclee@cc.nctu.edu.tw}}
\address{Department of Electrophysics, National Chiao-Tung University, Hsinchu,\\
30050,Taiwan}
\maketitle

\begin{abstract}
We perform the consistent quantization of open string D-brane in
non-constant NS-NS closed string B field background by directly imposing the
worldsheet conformal symmetry. In addition to the previous noncommutative
D-brane coordinates quantization with constant noncommutative parameter $%
\theta $, we obtain a set of constraints on the B field which can be
interpreted as a quantum consistent deformation of a curved noncommutative
D-brane.
\end{abstract}

\pacs{Noncommutative, D-brane.}

I. Introduction

\smallskip\ 

Historically, consistent quantization of string theory gave a remarkable
result that the embedding spacetime dimension is restricted to D = 26. It
was later realized in 80s that this is equivalent to requiring the conformal
symmetry on the worldsheet in the presence of target space flat metric $\eta 
$ background. For the last few years, it has been pointed out that when a
constant NS-NS B field background is turned on on the D-brane, consistent
quantization requires, in addition to D = 26, the noncommutativity of
spacetime coordinates on the D-brane\cite{1}. From the worldsheet conformal
symmetry point of view, this is a new exact ( all orders in $\alpha \acute{}$
) fixed point of the theory corresponding to flat metric $\eta $ and
constant B background. In the so-called Seiberg-Witten limit\cite{2},
Seiberg and Witten showed that all small perturbation calculation around
this new fixed point, including the effective field theory and correlation
functions etc. are dictated by replacing the ordinary operator product
expansion, which has been used in the ''old fixed point'', by the Moyal
star-product. For example, noncommutative Yang-Mills theory naturally
appears in the low energy description of D-brane with constant B background
in the Seiberg-Witten limit.

In this brief report, we consider the quantization of D-brane in the
non-constant but slowly varying B background. In this case, it is obvious
that the conformal symmetry or quantum consistency will be spoiled and one
needs to modify the quantization. There are two possible strategies to
restore the quantum consistency. One is to generalize the Moyal bracket to
more complicated types. The other one is to put constraint on the B
background but preserve the Moyal bracket quantization. Previous approach on
this subject has been based on the first method\cite{3}. The authors of \cite
{3} found that, in the case of vanishing field strength H = dB = 0, one get
the same noncommutative Moyal structure but with nonconstant $\theta $. With
the generic nonvanishing H $\neq $0 case, one gets a more general
noncommutative space with $\theta $ depending on X and P. However, their
treatment is valid only at the Poisson level, thus the quantum consistency
is still not guaranteed and it is not clear whether the resulting theory is
a consistent string theory. We will adopt the second approach in this paper
and impose the conformal symmetry on the worldsheet by directly calculating
the T $\cdot $T algebra and obtain a set of constraints on B field. In
contrast to the previous calculation which is valid only at Poisson level,
our result here is an exact consistent quantization and thus resulting a
consistent string theory.

The worldsheet action of an open string ending on a Dp-brane is

\begin{equation}
S=\frac{1}{4\pi \alpha \acute{}}\int_{\Sigma }d^{2}\sigma [g^{\alpha \beta
}\eta _{\mu \nu }\partial _{\alpha }X^{\mu }\partial _{\beta }X^{\nu
}+\epsilon ^{\alpha \beta }B_{\mu \nu }\partial _{\alpha }X^{\mu }\partial
_{\beta }X^{\nu }]+\frac{1}{2\pi \alpha \acute{}}\oint_{\partial \Sigma
}d\tau A_{i}\partial _{\tau }X^{i},
\end{equation}

where $A_{i},i=0,1,\cdot \cdot \cdot ,p,$ is the U(1) gauge field on the
Dp-brane. We will consider the case that both ends of the string are
attaching to the same Dp-brane, and B is turned on only on the Dp-brane. In
this case, the action in eq(1) can be rewritten as

\begin{equation}
S=\frac{1}{4\pi \alpha \acute{}}\int d^{2}\sigma [g^{\alpha \beta }\eta
_{\mu \nu }\partial _{\alpha }X^{\mu }\partial _{\beta }X^{\nu }+\epsilon
^{\alpha \beta }F_{ij}\partial _{\alpha }X^{i}\partial _{\beta }X^{j}],
\end{equation}

where F = B - dA. Note that F is invariant under the usual U(1) gauge
transformation of A and the following transformation

\begin{equation}
B\rightarrow B+d\Lambda ,A\rightarrow A+\Lambda .
\end{equation}

\smallskip

\smallskip II. Constant B Background

\smallskip

We first consider the case B = B$^{(0)}$ = constant and A = 0. The
worldsheet equation of motion of eq(2) with prescribed boundary conditions
can still be solved exactly as in the B = 0 case, and one can calculate the
energy momentum tensor to be

\begin{equation}
T_{++}=\frac{1}{2}\eta _{\mu \nu }\partial _{+}X^{\mu }\partial _{+}X^{\nu }=%
\frac{1}{2}M_{ij}\partial _{+}\widetilde{X^{i}}\partial _{+}\widetilde{X^{j}}%
+\frac{1}{2}\eta _{ab}\partial _{+}X^{a}\partial _{+}X^{b},
\end{equation}

where $a=p+1,\cdot \cdot \cdot ,9,$ and a similar formula for$T_{--}$. Note
that the region of worldsheet space coordinate of both $T$ is $0\leq \sigma
\leq \pi .$In equation (4), $\widetilde{X^{i}}\equiv X^{i}\mid _{F=0}$ and
the effective open string metric\cite{2} is

\begin{equation}
M_{ij}=\eta _{ij}-B_{ik}^{(0)}\eta ^{kl}B_{lj}^{(0)}.
\end{equation}

It is important to note that even in the presence of constant B$^{(0)}$
background, one still has the following open string continuing relation

\begin{equation}
X\acute{}(\sigma )=-X\acute{}(-\sigma )
\end{equation}

as in the case of B=0 background\cite{4}. This enables one to continue the
definition of $T_{++}$ to the region $-\pi \leq \sigma \leq \pi $ and to
eliminate $T_{--}$. The conformal algebra of the Virasoro operator $L_{n}$
can thus be, as in the case of B = 0 , translated to a single $T\cdot T$
algebra when one conformally maps the worldsheet from the cylinder to the
complex plane. Hereafter we will neglect the irrelevant second term in
equation (4) and refer $X^{i}$ to $\widetilde{X^{i}},T$ to the continuing $%
T_{++}.$

The quantization of the system was done by several groups\cite{1}. One way
to do it is to use an old formula of propagator of the action in eq (1) \cite
{5} ,and evaluate it at worldsheet boundary\cite{2},one gets

\begin{equation}
\langle X_{i}(\tau )X_{j}(\tau \acute{})\rangle =-M_{ij}\log (\tau -\tau 
\acute{})^{2}+\frac{i}{2}\theta _{ij}\epsilon (\tau -\tau \acute{}),
\end{equation}

where $\epsilon (\tau )$ is the function that is 1 or -1 for positive or
negative $\tau ,$and

\begin{equation}
\theta _{ij}=-2\pi (M^{-1}F)_{ij}.
\end{equation}

The original calculation \cite{5}of the full propagator was done by the
standard background field method with Riemann normal coordinate expansion,
and was considered to be a string-loop effect since there is no closed
string B field at open string-tree diagram without D-brane.With the
introduction of D-brane\cite{6} into the theory, the propagator becomes an
important string-tree effect.One remarkable observation of eq (7) is that
the D-brane coordinates become noncommutative

\begin{equation}
\lbrack X^{i}(\tau ),X^{i}(\tau )]=i\theta ^{ij}.
\end{equation}

This result is not surprising from quantum Hall theory point of view, where
the projective representation of magnetic translation group and the Moyal
realization of $W_{1+\infty }$symmetry \cite{7} were discussed some time
ago. We are now in a position to check the worldsheet conformal symmetry by
calculating the $T\cdot T$ algebra. The calculation in the bulk is the same
as in the free case. A direct calculation on the boundary gives the right
conformal algebra

\begin{equation}
T(\tau )\cdot T(\tau \acute{})\thicksim \frac{1}{2}\frac{M_{ij}M^{ij}}{(\tau
-\tau \acute{})^{4}}+\frac{2}{(\tau -\tau \acute{})^{2}}\frac{1}{2}%
M_{ij}\partial _{\tau \acute{}}X^{i}\partial _{\tau \acute{}}X^{j}+\frac{1}{%
(\tau -\tau \acute{})^{{}}}\partial _{\tau \acute{}}(\frac{1}{2}%
M_{ij}\partial _{\tau \acute{}}X^{i}\partial _{\tau \acute{}}X^{j})+\cdot
\cdot \cdot \cdot .
\end{equation}

Note that the first term on the r.h.s. is the same as in the B = 0 case, $%
M_{ij}M^{ij}=d=$spacetime dimension, and will be cancelled by the conformal
ghost. It is now clear that the propagator in eq (7) and eq (4) is a
consistent quantization. One notes that turning on the constant B$^{(0)}$
field means $\eta _{ij}\rightarrow M_{ij}$and thus simultaneously turning on
the noncommutative parameter $\theta .$

\smallskip

III. Non-constant B Background

We now turn to the non-constant B background.We will assume the weak or
slowly varying B field

\begin{equation}
B=B^{(0)}+b(X),
\end{equation}

where$B^{(0)}$ is a constant and $b(X)$ is a slowly varying field. In this
case, the worldsheet equation of motion of $X^{i}$can not be exactly solved
and eq (6) no longer holds. It turns out not possible for generic $b$ to
write down $T(\tau )$ with finite number of terms to close the $T\cdot T$
algebra. The treatment of \cite{3}was to modify eq (9) at Poisson level, but
they did not check the most important worldsheet conformal symmetry. To
avoid this difficulty, we will further assume$H\equiv db=0$ and trade $b$ to 
$dA$ by using eq (3). Since B is not turned on in the bulk, the $T\cdot T$
algebra is still valid there. We then propose the following boundary $T(\tau
)$

\begin{equation}
T(\tau )=\frac{1}{2}M_{ij}\partial _{\tau }X^{i}\partial _{\tau
}X^{j}+M_{ij}A^{i}\partial _{\tau }X^{j},
\end{equation}

where the open string metric $M_{ij}$ is due to $B^{(0)}$only.This form of $%
A^{i}$coupling is consistent with worldsheet vertex operator consideration.
We then check the worldsheet conformal symmetry by calculating the $T\cdot T$
algebra and neglect the second order $A^{i}$ term

\begin{center}
\begin{eqnarray}
T(\tau )\cdot T(\tau \acute{}) &\thicksim &\frac{1}{2}\frac{d}{(\tau -\tau 
\acute{})^{4}}+\frac{2}{(\tau -\tau \acute{})^{2}}(\frac{1}{2}M_{ij}\partial
_{\tau \acute{}}X^{i}\partial _{\tau \acute{}}X^{j}+M_{ij}A^{i}\partial
_{\tau \acute{}}X^{j})+  \nonumber \\
&&\frac{1}{(\tau -\tau \acute{})^{{}}}\partial _{\tau \acute{}}(\frac{1}{2}%
M_{ij}\partial _{\tau \acute{}}X^{i}\partial _{\tau \acute{}%
}X^{j}+M_{ij}A^{i}\partial _{\tau \acute{}}X^{j})+  \nonumber \\
&&\frac{2}{(\tau -\tau \acute{})^{3}}M_{ij}\partial ^{(j}A^{i)}+\frac{1}{%
(\tau -\tau \acute{})^{2}}M_{ij}\partial ^{i}\partial ^{j}A_{k}\partial
_{\tau \acute{}}X^{k}.
\end{eqnarray}
\end{center}

The worldsheet conformal symmetry requires the following constraints on the
background field A

\begin{equation}
M_{ij}\partial ^{(j}A^{i)}=0,M_{ij}\partial ^{i}\partial ^{j}A_{k}=0.
\end{equation}
The constraints on the background A can then be traded back to the
background $b$ to be

\begin{equation}
M_{ij}\partial ^{j}b^{ik}=0,M_{ij}\partial ^{i}\partial ^{j}b_{kl}=0.
\end{equation}

In conclusion, the consistent quantization of D-brane in the presence of
non-constant, slowly-varying background B in eq (11) with vanishing field
strength $H\equiv dB=0$ requires constraints in eq (15) on the B field.
Equation (14) is consistent with the equation of motion for noncommutative
Yang-Mills in the weak field approximation. Equation (15) can be interpreted
as a quantum consistent deformation of noncommutative D-brane by background
B field, and is presumably related to a {\it curved noncommutative D-brane}
characteristized by a X dependent $\theta (X)$. This is consistent with
ref[3] although our treatment here is beyond the Poisson level. One expects
that a modified associated star-algebra can be constructed on such a new
curved noncommutative D-brane. It seems that for the more generic H $\neq $0
case, one needs to introduce a nonassocitaed star-algebra to incorporate the
deformation. On the other hand, in this  H $\neq $0 case one is forced to
take an infinite number of terms on the r.h.s. of either eq (9)\cite{3} or
eq (12) to close the conformal algebra, and get a consistent string theory.
This is reminiscent of the case of closed string with closed string
background around the ''old fixed point'', where the corresponding 2d $%
\sigma -$ model was shown to be nonperturbative nonrenormalizable in the
weak field expansion, and an infinite number of counter terms was needed to
preserve the worldsheet conformal symmetry.\cite{8}.

\smallskip

Acknowledgement

I would like to thank Pei-Ming Ho for many discussions. This research is
supported by National Science Council of Taiwan, R.O.C., under grant number
NSC 89-2112-M-009-045.The support from string study group of NCTS of Taiwan
is also acknowledged.

\end{document}